\documentclass[12pt,amsmath,amssymb,prb]{revtex4}
\usepackage{amsmath}
\usepackage{amsfonts}
\usepackage{amssymb}
\usepackage{graphicx}
\usepackage{varioref}
\ifx\pdfoutput\undefined
\usepackage[unicode,dvips]{hyperref}
\else
\usepackage[bookmarks=false, colorlinks, unicode, pdfstartview=FitH]{hyperref}
\fi
\usepackage{afterpage}
\usepackage{ifthen}
\usepackage{hyperref}

\begin{document}
\title{Quadratic Magnetoelectric Effect and Magnetic Field Induced Pyroelectric Effect in Multiferroic $BaMnF_4$}
\author{À.K. Zvezdin, G.P. Vorob'ev, A.M. Kadomtseva, Yu.F. Popov, D.V. Belov, A.P. Pyatakov}
\affiliation{A.M. Prokhorov General Physics Institute RAS, Vavilova 38, Moscow, 119991}
\affiliation{M.V. Lomonosov Moscow State University, Leninskie gori, MSU, Moscow, 119992}

\begin{abstract}
The experimental studies of magnetoelectric effects in pulse
magnetic field up to 250~kOe and their theoretical analysis on the
basis of magnetic symmetry consideration are carried out. It is
shown that the nonvanishing components of quadratic magnetoelectric
effect tensor corresponding to the electric polarization along b-
and ñ-axes point out the triclinic distortion of the crystal
symmetry. Anomalous temperature dependence of magnetically induced
polarization $P_a(H_b)$  testifies to the magnetically induced
pyroelectric effect. The torque curves measurements show the
deflection of the spin orientation from the b-axis at 9 degrees of
arc.

The work is supported by
 "Progetto Lagrange-Fondazione CRT" and RFBR (07-02-92183-NCNI-a).
\end{abstract}

\maketitle \sloppy \noindent{\bf }

The fluoride $BaMnF_4$ is a classical ferroelectromagnet (in modern terms
multiferroic) and it has been the subject of intensive research for
some decades (see, for example, the review [1] and reference
therein, and also later publications [2-10]). It belongs to the
family of isomorphic compounds of general formula $Ba M F_4 $, where
$M = Mn, Fe, Co, Ni, Mg, Zn$, which are paramagnetic at room
temperature.
 Their crystal structure is described by the orthorhombic space group $A2_1am $ (a=5.9845, b =
15.098, c = 4.2216) in international notation (or $C_{2v}^{12}$ in
Schoenflies notation).  The $A2_1am $ group includes twofold $2_1$ screw
axis parallel to the $a$-axis with the shift of $a/2$ , the glide
symmetry plane $a$ perpendicular to $b$-axis with the shift along
 $a$-axis at $a/2$, two mirror planes $m$ perpendicular to the $c$-axis, located at $z= 0, c/2$  and corresponding translations.
 In contrast to the other fluorides the $BaMnF_4$ is pyroelectric, besides the ferroelectric and antiferromagnetic
 phase transition ($T_C \approx 1113~K, T_N \approx 25~K$ [10-12]) it has structural phase
transition at $T_0 \approx 250~K$ into incommensurate monoclinic
phase, that retains at low temperatures.  In $BaMnF_4$ spontaneous
electric polarization
 $\bold{P}_s  \parallel \bold{a}$. Symmetry allows the existence of antiferroelectric ordering along $b$-axis,
but forbids one along $c$-axis.

Structural phase transition into the incommensurate phase at $T_0=
250~K$ in $BaMnF_4$, is described by the wave vector $\bold {q}=
(\mu /a, 0.5,0.5)$, where $\mu = 0.39$. That means the size of the
primitive cell doubles in the $bc$-plane and incommensurate structure along
the a-axis with a period $a/\mu=15\AA$ occurs. It is established, that this
transition at $T_0= 250K$ is ferroelastic one [2,3].

Magnetoelectric properties of the $BaMnF_4$ are still not completely
understood. For example the symmetry allows linear magnetoelectric
effect and ferroelectrically induced weak ferromagnetism [1] but they
are not observed experimentally (presumably their volume averaged
values are zeroed out due to the existence of incommensurate
structure). Quadratic magnetoelectric effect though observed [2] was
measured only for polarization along \emph{à}-axis, while the
measurements of other components can serve as sensitive indicator of
symmetry distortions. As in $BaMnF_4$ ferroelectric and
ferromagnetic ordering coexist, the heat induced pyroelectric and
magnetocaloric effects, as well as magnetoelectric effect, mediated
by heating, are possible.

In this paper the magnetoelectric dependencies in high magnetic
field up to 250 kOe at various direction of polarization and
magnetic field, as well as torque curves in static fields up to 12
kOe are measured for the first time in single crystal $BaMnF_4$ in
the temperature range 4.5-150 Ê. They are interpreted on the basis
of crystal and magnetic symmetry as well as magnetic structure and
magnetic field induced phase transitions. The anomalous temperature
dependence of magnetically induced polarization $P_a(H_b)$  is
explained by magnetocaloric and pyroelectric effects.

\noindent\textbf{1.Experiment }
    The magnetic field induced electric polarization measurements showed that longitudinal quadratic magnetoelectric effect
    is maximum along a-axis of the crystal ($\beta_{11}\approx 1.6\cdot
    10^{-19}sA^{-1}$ at $4.5~K$), along which spontaneous polarization is directed (fig. 1). The transversal electric polarization
    $P_b(H_a)$ was also proportional to the square of magnetic field ($\beta_{21}\approx 1.4\cdot
    10^{-20}sA^{-1}$ at $30~K$), lowering with the growing temperature (fig. 2).
    The longitudinal electric polarization dependence $P_c(H_c)$ (fig. 3) has somewhat intricate character and differers qualitatively from $P_a(H_a)$.
    At temperature 4.5 K there was maximum of negative electric polarization in the field $\sim $50 kOe, followed by decrease of absolute value
    and sign reversal at 200 kOe. At higher temperatures up to $T_N=25~K$ the character of the dependencies $P_c(H_c)$ was similar,
    and maximum of electric polarization shifted in the region of higher fields $\sim $ 100 kOe.
    Complicated character of $P_c(H_c)$ dependence points out the transformation of magnetoelectric interactions that
    possibly related to the magnetic structure changes in high magnetic field,
    particularly due to the spin modulated structure suppression as it was observed in $BiFeO_3$ [13].
    In the magnetic field oriented along \emph{b}-axis the anomalies of magnetoelectric effect at spin-flop phase transition were observed ($H_{SF}\approx 10 $kOe).
    In fig. 4 the magnetic field dependencies of the longitudinal magnetoelectric effect $P_b(H_b)$ are shown.
    The jump of electric polarization at $H\sim 10$ kOe is followed by quadratic magnetoelectric dependence,
    that is related to the change of the symmetry from 2' to 2. At H||b along \emph{à}- and \emph{ñ}- axes (fig. 5, 6)
    the jumps of electric polarization at H$\approx $ 10 kOe were observed. The transversal electric polarization
    along the c-axis increased with temperature decrease, while the one along the $a$-axis decreased, vanishing at Ò=4,5K.

    The torque curves in the bc-plane are shown in figure 7. Below Neel temperature the distinct anomalies (jumps) at magnetic field orientation
    $90\pm9^{\circ}$ to $b$-axis ($\varphi \approx 150^{\circ}$ in the figure) take place.

\noindent\textbf{2. Discussion.} In the high symmetry orthorhombic phase
($2mm$) of $BaMnF_4$ the linear magnetoelectric effect is apparently absent,
and the quadratic effect is expressed by formula
\begin{equation}\label{1}
P_i = \beta_{ijk} H_j H_k,
 \tag{1}
\end{equation}
where the magnetoelectric tensor of the 3-rd rank is [14]

\begin{equation}\label{2}
   \begin{pmatrix}
     \beta_{11} & \beta_{12} & \beta_{13} & 0 & 0 & 0 \\
    0 & 0 & 0 & 0 & 0 & \beta_{26} \\
    0 & 0 & 0 & 0 & \beta_{35} & 0 \
   \end{pmatrix}.
   \tag{2}
  \end{equation}
The coefficients $\beta_{nm}$ are related to the 3-rd rank tensor
components by standard rule:
$$
\beta_{in} = \beta_{ijk}  \, (jk \rightarrow n= 1,2, ..., 6).
$$
It is believed that the structural phase transition at $T=250~K$ is the
transition to the incommensurate improper ferroelastic phase with
the averaged monoclinic symmetry ($P2_1$)[2, 3]. The distortion of
original orthorhombic structure in the low temperature phase is small
and can be characterized by an angle $\alpha \backsim 10^{-2} [2]$.
It is natural to assume the angle $\alpha$ as a small parameter of theory,
characterizing the change of physical values at the phase transition
from orthorhombic to monoclinic phase.

In monoclinic phase the components of magnetoelectric tensor
$\beta_{14}, \beta_{25},  \beta_{36} $ become nonvanishing.
Obviously, their values should be proportional to $\alpha$.
Similarly, the corresponding changes of nonzero components (2) are
also proportional to $\alpha$,  i.e.
$\Delta{\beta_{11}}/{\beta_{11}} \backsim \alpha $  etc.

In accordance with [2], in the $BaMnF_4$ \, $\beta_{11} =
1.1\cdot10^{-19}sA^{-1}$, $\beta_{13} = -1.6\cdot10^{-19}sA^{-1}$,
that agrees well with the results of our measurements ($\beta_{11} =
1.6\div 0.8\cdot 10^{-19}sA^{-1}$ in the temperature range $4.2\div
15~K$) and is close to the corresponding coefficients in $BiFeO_3$
($\backsim{10^{-19}sA^{-1}}$ at 4.2 K [15]), though somewhat lower
than in $PbFe_{0.5} Nb_{0.5}O_3$ ($10^{-17}-10^{-18} sA^{-1}$ at
15~Ê)[16] and $NiSO_4\cdot6H_2O$, ($10^{-17}sA^{-1}$ at $4.2Ê$
[17]). There are no data about other components of $\beta_{ijk}$  in
$BaMnF_4$ in literature. Authors [2] carried out the measurements of the
quadratic magnetoelectric effect of $BaMnF_4$ in the crystal planes
$ac$ and $bc$, the jump of electric polarization of unknown nature
was observed that, probably, hindered from determining
$\beta_{xyy}$ and $\beta_{xyz}$.

It is worth noticing that according to [2], in crystal with monoclinic symmetry
(class $2$, with 2-fold axis along a-axis) the components of electric polarization
$p_y$  and $p_z$ should be zero if external field is directed along
crystal axes. However the group $2$, assumed in [1-4]
 as an averaged symmetry of the $BaMnF_4$ at  $T<T_0$, is somewhat approximated.
Indeed, as the crystallographic structure in the $BaMnF_4$ at $T<T_0$ is incommensurate along a-axis, the element
$2$ is obviously violated [1]. Strictly speaking the averaged
symmetry should lower to triclinic, in which all elements of the matrix
(2) are nonvanishing [1].  Of course the difference of the "triclinic"
tensor $\beta_{ijk}$ from the "monoclinic" (2) in this case is minor with the small parameter $\thicksim
(a/\lambda)\alpha^{1/2} $, where $\lambda$ is the wavelength of incommensurate modulation, $a$ is the lattice constant along the $a$-axis.

Let us consider from this point of view the experimental data
(Fig.2,4,6). Indeed, at $T<T_0$ the measured values of the components
$\beta_{21 }, \beta_{24}, \beta_{34} $ are at least an order smaller
than $\beta_{11}$ (Fig.1)  that is naturally explained by the
smallness of triclinic distortion.

At $T<T_N$  the new features arise in magnetoelectric behavior
related to the magnetic order and transformation of magnetic
structure in external field. Let us consider in more detail the
ground state of the crystal in low symmetry phases in terms of
magnetic and antiferromagnetic vectors $M$ and $L$.
Neutron diffraction studies provide with contradictory information
about the antiferromagnetic structure $L$. According to [10]
vector \textbf{$L$}$\| b$-axis, while in [11,12] the L-vector lies
in $bc$-plane at the angle $\sim 9^{\circ}$ to $b$-axis.

Our measurements of torque curves (fig.7) testify to the deflection
from b-axis, the angles at which the anomalies (jumps) observed
 are shifted from the middle position at the angles $\pm9^{\circ}$.
These jumps can not be explained by $180^{\circ}$ reorientation of
weak ferromagnetic moment $M$ along c-axis that was introduced in [1,12,18],
 otherwise the weak ferromagnetic moment would reorient from
parallel to antiparallel position with respect to the magnetic field
that is unlikely. Qualitatively the torque curves at $H<H_{SF}$
can be explained  by existence of four (effectively two) phases with
the antiferromagnetic vector direction at $\pm9^{\circ}$ to $b$-axis
and phase transition between them. However the quantitative description
of the torque curves requires taking into account the
antiferromagnetic domains that is beyond the scope of this paper.

Thus the magnetic structure in the ground state takes the form:
$\bf{L}=\left(0,\,L_{y},\,L_{z}\right)$, where
$\left({L_{z}}/{L_{y}}\right) = \tan{\psi}$, where $\psi = 9^o$, $\bf{M}\approx 0$.
This small deflection from b-axis can be interpreted as the consequence of "monoclinic"
contribution $K_{yz}L_yL_z$ into anysotropy energy. The value $K_{yz}$ is, obviously,
proportional to the monoclinic distorsion of the structure at $T<T_0$, i.e. $|K_{yz}/K_{rhomb}| \sim \alpha$,
where $K_{rhomb}$ is characteristic value of "orthorhombic" contribution to the magnetic anysotropy.

In magnetic field the transformation of antiferromagnetic structure occurs, i.e. spin-flip and spin-flop transitions
(depending on the direction of external magnetic field with respect to original orientation of antiferromagnetic vector).
In zero approximation on parameter $\alpha$ the ground state magnetic structures can be presented by formulae:
$$ \mathbf{H} = (H,0,0)$$ $$  \mathbf{M}= (\chi_\bot H, 0,0),
\mathbf{L}=(0,L_y,0),$$

$$\mathbf{H} =(0,H,0)$$ $$  \mathbf{M}= (0,\chi_\| H,
0),\mathbf{L}=(0,L_y,0),  H<H_{SF},$$ $$\mathbf{M}= (0,\chi_\bot H,
0),\mathbf{L}=(0,0,L_z),  H>H_{SF},$$

$$\mathbf{H} = (H,0,0)$$ $$  \mathbf{M}= (\chi_\bot H, 0,0),
\mathbf{L}=(0,L_y,0), ),$$ where $\mathbf{M}$ is the magnetization,
$\chi_\bot, \chi_\|$ are perpendicular and parallel susceptibility of
the material, $H_{SF}$ is the spin flop field ($H_{SF} \sim 10^4 $
Oe in $BaMnF_4$ [2]). Taking into account the aforementioned
monoclinic contribution into magnetic anisotropy would bring us out of
the zero approximation on $\alpha$ and would lead to more bulky formulae
but would not change the qualitative picture of the phenomenon.

In those cases when magnetic field is directed along $a$ and $c$ -
axes of the crystal, the magnetic field induced components of electric polarization
$\Delta p_i, i=x,y,z$ can be described on the basis of the same simple symmetry consideration
as in (1):
\begin{equation}\label{4}
\Delta p_i = a_i M_i^2 + b_iL_y^2 \sim H^2,
\tag{4}
  \end{equation}

where $a_i$, $b_i$, are the coefficients of series expansion. It agrees
qualitatively with the behaviour of the magnetoelectric curves in figures 1-3.

The different situation is at $\bold{H} = (0,H,0)$. The
spin flop in $Mn$-ions system is accompanied by the jumps
of antiferromagnetic vector components and magnetization. The value
of the jump is a maximum at $T\rightarrow 0 K$ and tends to zero 0 at
$T\rightarrow T_N$, where $T_N$ is Neel temperature. According to
(4), the jumps of electric polarization should copy this behavior.
At figures 4 and 6 the jumps of $\Delta p_y$, $\Delta p_z$ are
clearly seen and their temperature dependences agree qualitatively
to the described scheme.

However, the  electric polarization $\Delta p_x$ demonstrates
counter intuitive behaviour (fig. 5). At $T\rightarrow 0~K, H\|b$
the jump $\Delta p_x$ is vanishing and increase at $T\rightarrow
T_N$. Note also that absolute values of the electric polarization
$\Delta p_x$ is substantially larger than $y$ and $z$ components.

The probable reason for such an unusual behavior of $p_x(H)$ can be
the result of pyroelectric effect caused by magnetocaloric heating
of the $BaMnF_4$ sample in pulse magnetic field.
 Indeed, the pyroeffect is determined in this case:
$$ \delta p_x = P_0 \ c \  \delta T ,$$
where $P_0$ is the value of spontaneous electric polarization, $c$
is the constant, that depends on crystallographic parameters of the
lattice and thermal expansion tensor components (see  [19], for
example). The value $\delta T$ is determined by magnetocaloric
effect:
$$ \delta T = c_H^{-1}  \frac{\partial M_H}{\partial T}\Delta H,  $$
where $c_H$ is the heat capacity at constant field, $M_H$ is the projection
of the magnetization on magnetic field direction, $\Delta H$ is the magnetic field increment.
In figure 8 the temperature dependence of $\frac{\partial M_H}{\partial T}$ is shown,
that qualitatively corresponds to the measured anomalies in $p_x(H,T)$ at $T<T_N$.

It is worth noticing that mentioned mechanisms do not exhaust the
whole range of possibilities of magnetically induced electric
polarization appearance in $BaMnF_4$, that possess incommensurate
ferroelastic structure. The full picture includes piezoelectric
effect induced by magnetostriction, and flexoelectric response, that
appears in spatially modulated incommensurate structure under the
influence of magnetostrictive deformations.

The first one, the piezoelectric response, is determined as $$\delta P_{\alpha}^{piezo} =
C_{\alpha\beta\gamma} \ \varepsilon_{\beta\gamma}(H),$$
 where $\varepsilon_{\beta\gamma}$ is the tensor of magnetostriction deformation, $ C_{\alpha\beta\gamma}$
 is the tensor of piezoelectric coefficients.

 The flexoelectric response is determined by formula [19]: $$ \delta P_\alpha^{flexo} = f_{\alpha\beta\gamma\delta}
  \frac{\partial^2 r_\beta (H)}{\partial x_\gamma\partial x_\delta}, $$ where $\frac{\partial^2 r_\beta}{\partial x_\gamma\partial_\delta}$
is the derivative of distortion tensor (the existence of
incommensurate structure in the $BaMnF_4$ evidences strogly the nonvanishing of the derivative), $f_{\alpha\beta\gamma\delta}$ is the tensor of flexoelectric coefficients.

We do not aim to make the detailed analysis of the
later mechanisms; its contribution to the observed
polarization should be smaller than quadratic magnetoelectric and
pyroelectric effects, but they can be responsible for irregularities
of magnetoelectric dependencies (Fig. 2,3).

\newpage
\begin{figure}[]
\includegraphics[width=7 cm,clip=true]{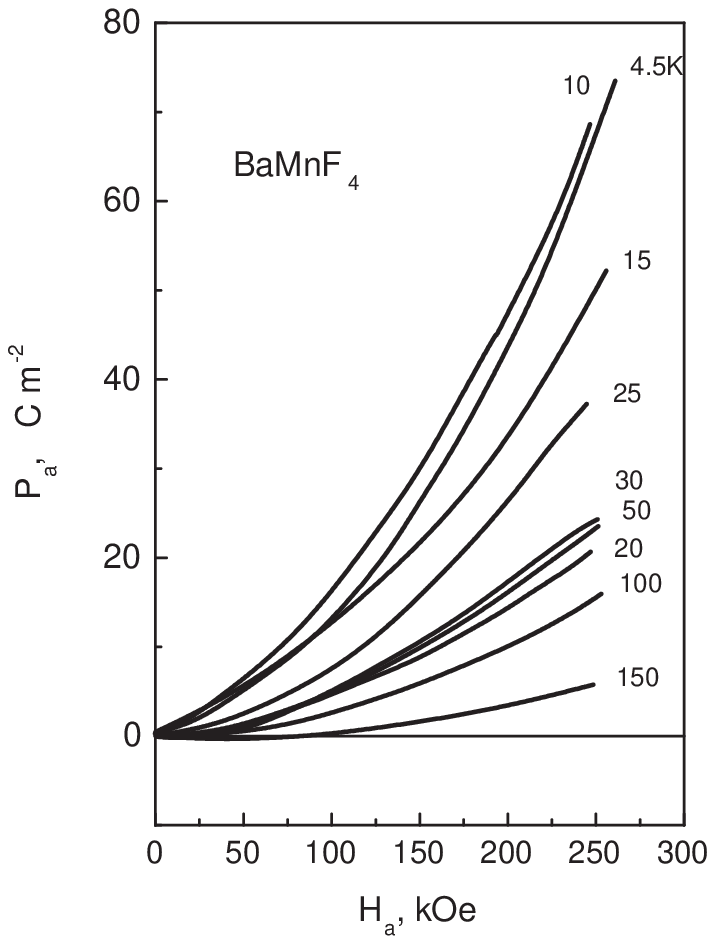}
\caption {Magnetoelectric dependencies $P_a(H_a)$ in pulse field at
various temperatures}
\end{figure}

\begin{figure}[]
\includegraphics[width=7 cm,clip=true]{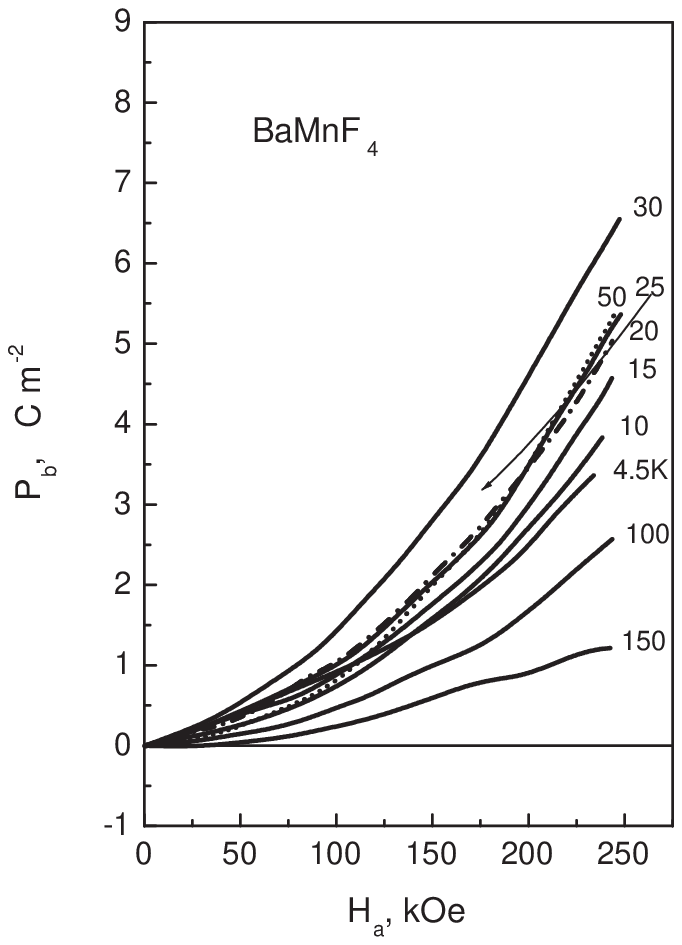}
\caption {Magnetoelectric dependencies $P_b(H_a)$ in pulse field at
various temperatures}
\end{figure}

\newpage
\begin{figure}[]
\includegraphics[width=7 cm,clip=true]{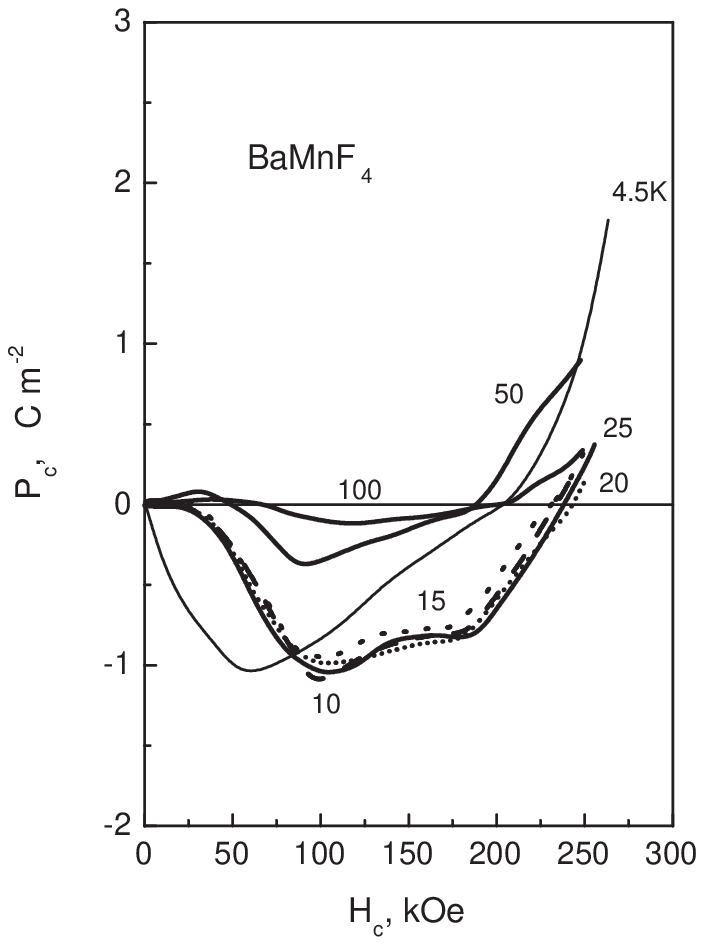}
\caption {Magnetoelectric dependencies $P_c(H_c)$ in pulse field at
various temperatures}
\end{figure}

\begin{figure}[]
\includegraphics[width=7 cm,clip=true]{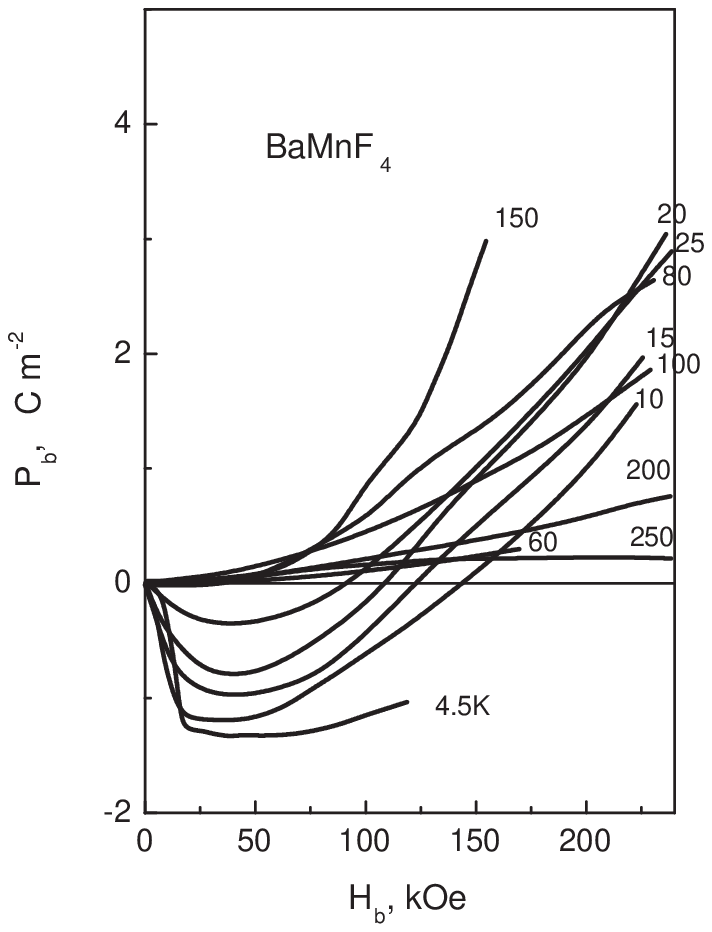}
\caption { Magnetoelectric dependencies $P_b(H_b)$ in pulse field at
various temperatures}
\end{figure}

\newpage
\begin{figure}[]
\includegraphics[width=7 cm,clip=true]{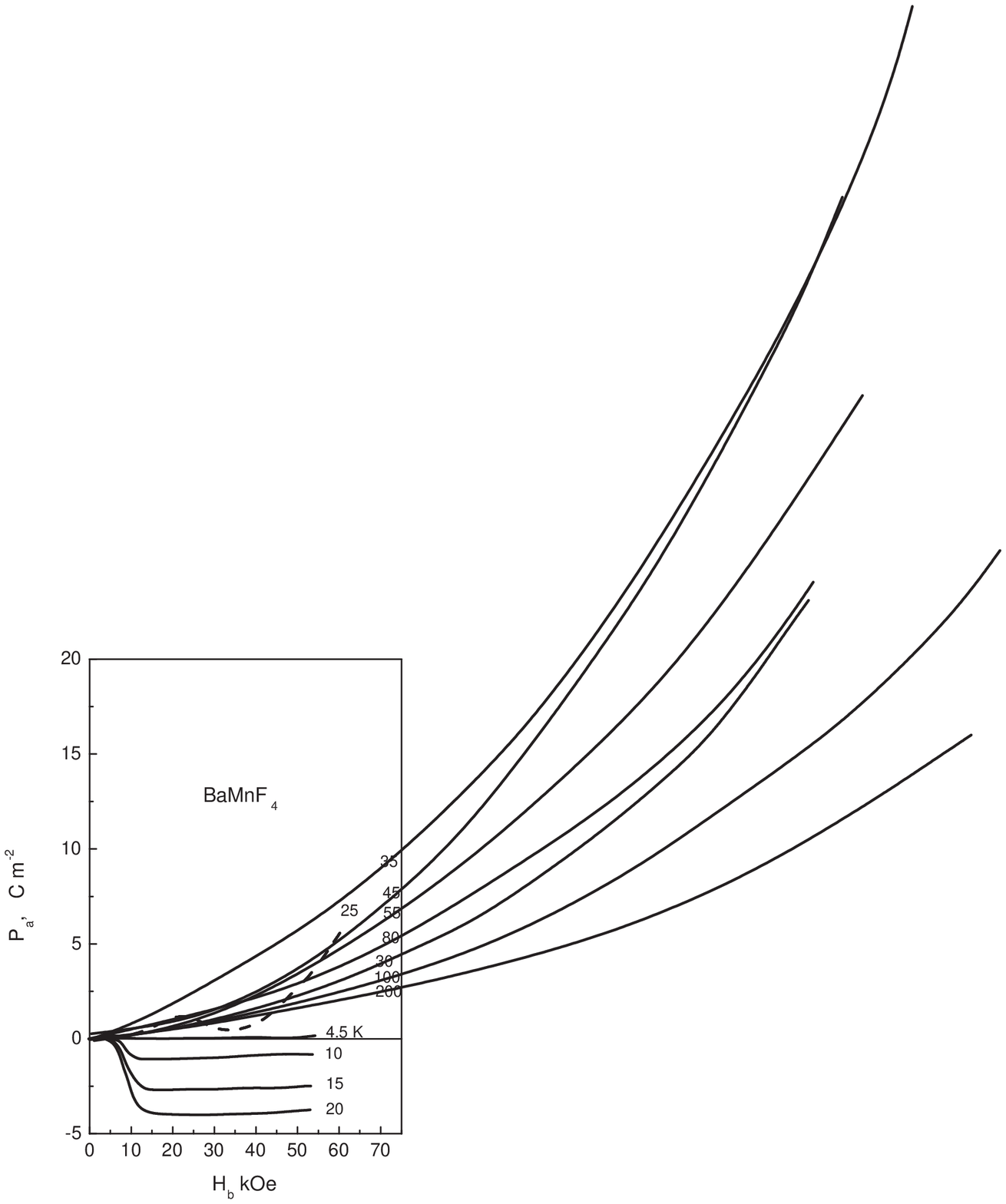}
\caption{Magnetoelectric dependencies $P_a(H_b)$ in pulse field at
various temperatures}
\end{figure}

\begin{figure}[]
\includegraphics[width=7 cm,clip=true]{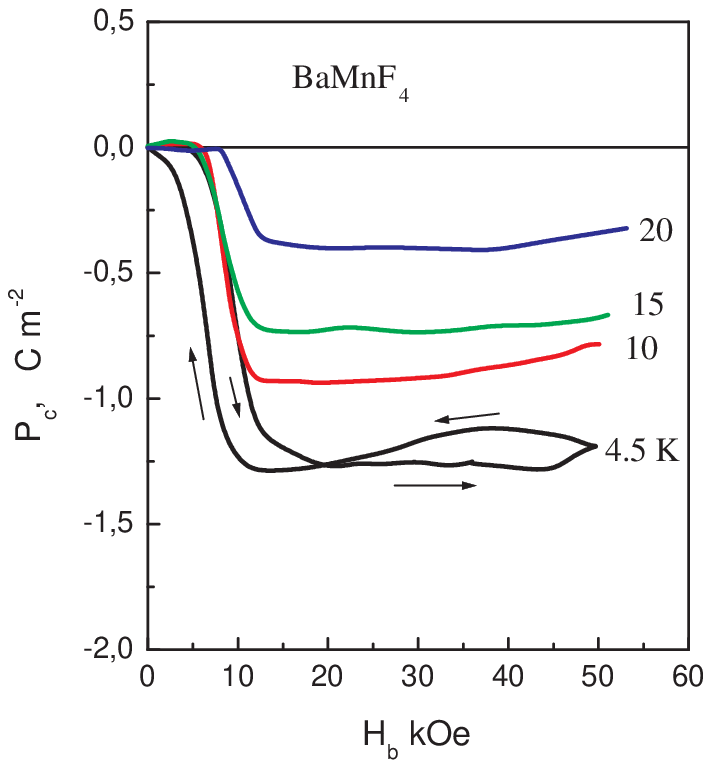}
\caption { Magnetoelectric dependencies $P_c(H_b)$ in pulse field at
various temperatures}
\end{figure}

\newpage
\begin{figure}[]
\includegraphics[width=10 cm,clip=true]{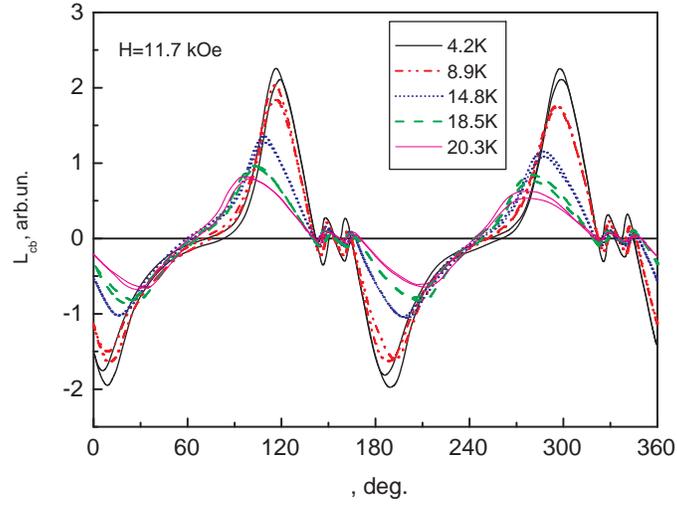}
\caption  {Torque curves at various temperatures in static magnetic
field $H<H_{spin flop}$}
\end{figure}

\begin{figure}[]
\includegraphics[width=10 cm,clip=true]{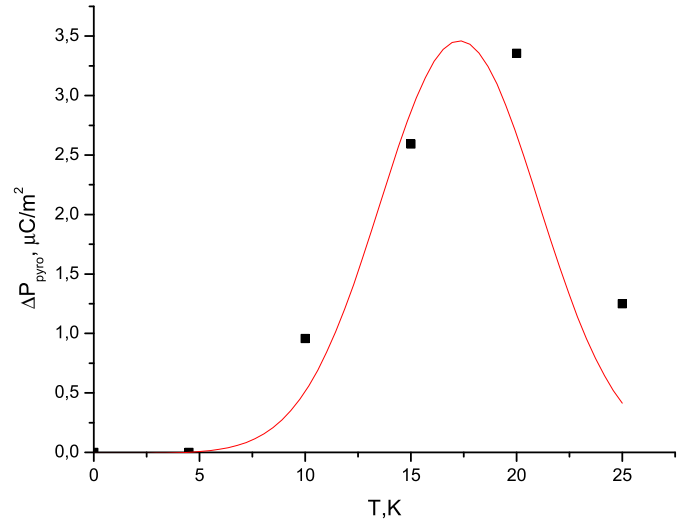}
\caption  {The schematic curve $ \frac{\partial M_H}{\partial T}=
\frac{\partial \chi(T)}{\partial T} \Delta H $ (line), and
magnetoelectric anomalies (points)}
\end{figure}

\end{document}